\documentclass{article}
\usepackage[utf8]{inputenc} 
\usepackage[T1]{fontenc}    

\usepackage{authblk}
\usepackage{graphicx}
\usepackage[square,numbers]{natbib}
\usepackage{doi}

\begin{document}

\title{Implantable Photonic Neural Probes with Out-of-Plane Focusing Grating Emitters}
\author[1,2,*] {Tianyuan Xue}
\author[1] {Andrei Stalmashonak}
\author[1,2] {Fu-Der Chen}
\author[1,2] {Peisheng Ding}
\author[3] {Xianshu Luo}
\author[3] {Hongyao Chua}
\author[3] {Guo-Qiang Lo}
\author[1] {Wesley D. Sacher}
\author[1,2,**] {Joyce K. S. Poon}

\affil[1]{Department of Nanophotonics, Integration, and Neural Technology, Max Planck Institute of Microstructure Physics, Weinberg 2, 06120 Halle, Germany}
\affil[2]{The Edward S. Rogers Sr. Department of Electrical and Computer Engineering, University of Toronto, 10 King's College Road, Toronto, Ontario M5S 3G4, Canada}
\affil[3]{Advanced Micro Foundry Pte. Ltd., 11 Science Park Road, 117685, Singapore}

\affil[*]{xuetiany@mail.utoronto.ca}
\affil[**]{joyce.poon@mpi-halle.mpg.de}
\date{January 10, 2024}

\begin{abstract}

We have designed, fabricated, and characterized implantable silicon neural probes with nanophotonic grating emitters that focus the emitted light at a specified distance above the surface of the probe for spatially precise optogenetic targeting of neurons. Using the holographic principle, we designed gratings for wavelengths of 488 and 594 nm, targeting the excitation spectra of the optogenetic actuators Channelrhodopsin-2 and Chrimson, respectively. The measured optical emission pattern of these emitters in non-scattering medium and tissue matched well with simulations. To our knowledge, this is the first report of focused spots with the size scale of a neuron soma in brain tissue formed from  implantable neural probes.

\end{abstract}

\maketitle

\section*{Introduction}

Genetically encoded optogenetic actuators enable the functional interrogation of complex neural circuits by providing a mechanism for the precise manipulation of neuronal activity with light \cite{Deisseroth2015}. The excitation spectra of optogenetic actuators, such as channelrhodopsin-2 (ChR-2), often lie in the visible wavelength range \cite{Lin2011,Klapoetke2014}. However, the attenuation length of light at these wavelengths in brain tissue is limited to <1 mm \cite{Al-Juboori2013,Adesnik2021}. Implantable solutions, such as optical fibers and implantable neural probes, can deliver illumination directly to deep brain regions beyond the attenuation limit.\cite{Neutens2023,Mohanty2020,Pisanello2018,Sacher2019,Sacher2021,Sacher2022,Chen2023}.

Implantable silicon (Si) neural probes leverage the dense integration of photonic and electronic circuits on Si to enable concurrent electrophysiology recording and optogenetic stimulation while maintaining a volume comparable to or smaller than that of other implantable approaches \cite{Neutens2023,Voroslakos2022,Mohanty2020,Sacher2021,Schwaerzle2017,Pisanello2018,Libbrecht2018}. 
While both {\textmu}LEDs and integrated photonic waveguide gratings have been used as light emitters on implantable Si probes \cite{Mohanty2020,Neutens2023,Kim2013,Sacher2019,Sacher2021,Chen2023}, grating emitters have several advantages compared to {\textmu}LEDs. Grating emitters do not generate heat aside from light absorption by brain tissue, whereas the low wall-plug efficiencies of {\textmu}LED emitters require mitigation of heating effects \cite{Voroslakos2022,Yasunaga2022,Kim2013}. Furthermore, because light scattering in tissue is highly directional, beam forming can be achieved through the design of gratings and optical phased arrays. To this end, we have previously demonstrated the emission of highly directional beams \cite{Sacher2019,Mu2023,Chen2023}, steerable directional beams \cite{Chen2021,Sacher2022,Sharma2023}, and light sheet beams \cite{Sacher2021,Chen2023} from grating emitters on implantable Si probes.

In these previous works, the intensity of light decayed monotonically away from the grating emitter, and neurons close to the surface of the probe were preferentially excited. However, tissue near the probe surface is also the most prone to damage by the implant \cite{Fiáth2019}. In this work, we have designed out-of-plane focusing grating emitters that focus the emitted light at a point above the surface of the neural probe for spatially precise targeting of neurons at a distance. The focusing of light emission has the additional benefit of reaching the required intensities for optogenetic actuation of $\sim$ 1 mW/mm\textsuperscript{2} \cite{Lin2011} at lower input powers compared to other emitters. This type of grating emitters has previously been designed for ion control \cite{Mehta2017,Corsetti2023}, memory addressing \cite{Becker2020} and neural probes  \cite{Lanzio2018}. However, in contrast to Lanzio et al. \cite{Lanzio2018}, here, the probes have been fabricated in a foundry and the optical emission pattern has been characterized in tissue. Our implantable neural probes contained up to 16 focusing grating emitters on shanks that were 6 mm long. To characterize the optical profile of these emitters, we captured the side-view beam profiles in a fluorescent dye solution and three-dimensional (3D) profiles using fluorescent photoresist in a water chamber. Lastly, we observed focusing of the emitted light in fixed brain tissue with genetically encoded calcium indicator (GECI) expression. To our knowledge, this is the first report of focusing of light emitted by a Si probe implanted in brain tissue. 

\section*{Design, Fabrication and Packaging}

\subsection*{Design Methodology}

\begin{figure}
    \centering
    \includegraphics[width=1\textwidth]{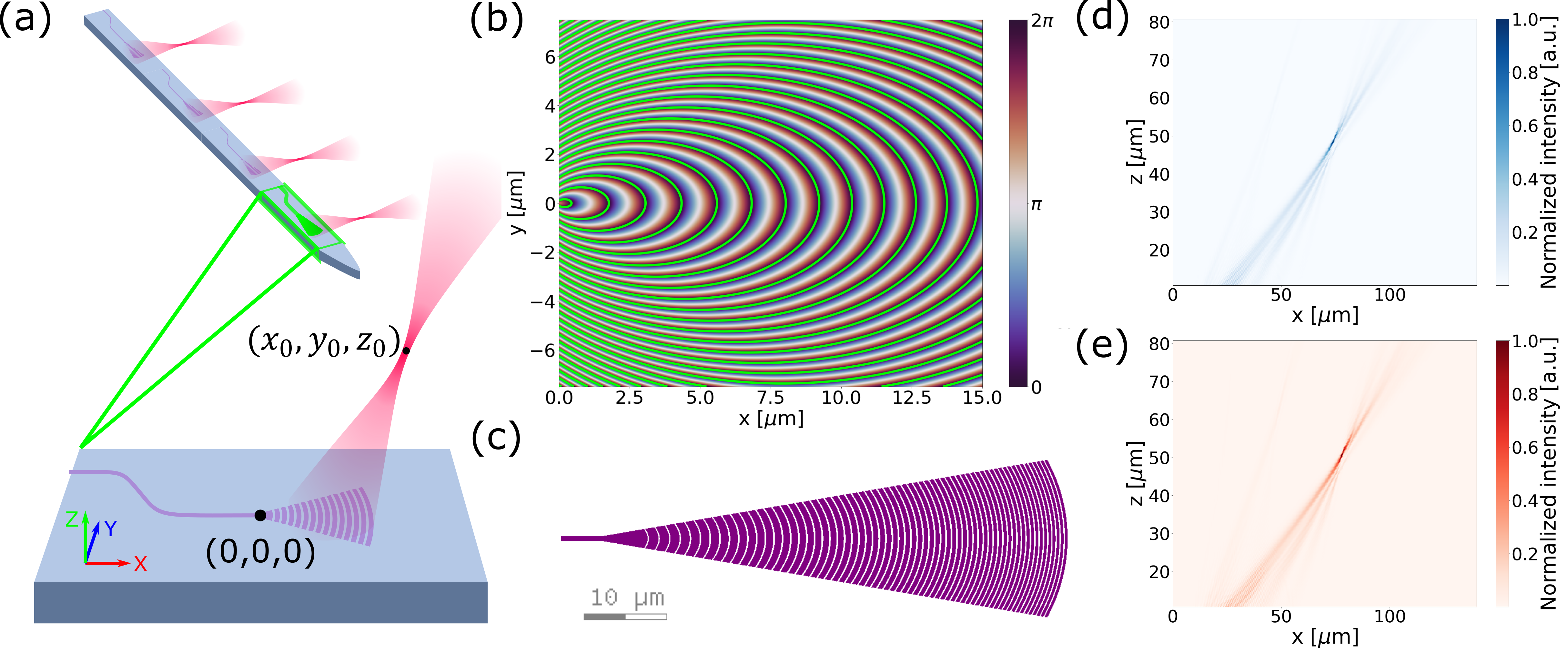}
            \caption{Grating emitter design. (a) Schematic overview of an out-of-plane focusing grating emitter with focal point located at coordinates ($x_0,y_0,z_0$). (b) Phase map generated for $\lambda$ = 488 nm. The contour lines dictated by the phase matching condition are overlaid in green. (c) The finalized layout design of the grating emitter for $\lambda$ = 488 nm. Emitted beam profile on the $y=0$ plane simulated in 3D FDTD for (d): blue emitter ($\lambda$=488 nm), and (e): red ($\lambda$=594 nm) emitter. }
            \label{fig:design}
\end{figure}

The probe was fabricated on a silicon (Si) substrate with a low-loss visible silicon nitride (SiN) waveguide layer on the platform  detailed in Chen et al. \cite{Chen2023}. To transform the incident wave to the desired output beam, we used the holographic principle to determine the set of grating curves to shape the output phase profile. Specifically, the curves are the 2$\pi$-spaced contours of the phase map resultant from the sum of the phases of the input and desired output waves. This is a modified version of the phase-matching condition found in Oton \cite{Oton2016}. In our phase matching condition, the incident and output phase profiles are prescribed as radial and spherical phase fronts, respectively, and are given as
\begin{equation}
2q\pi = n_{eff}k_0\sqrt{x^2+y^2}+n_{tissue}k_0\sqrt{(x-x_0)^2+(y-y_0)^2+z_0^2},
\end{equation}
where $(x_0,y_0,z_0)$ are the spatial coordinates of the intended focus, $n_{eff}$ and $n_{tissue}$ are the effective indices of the SiN grating and brain tissue respectively. Grating teeth defined using this phase-matching condition, as shown in Fig. \ref{fig:design} (b), results in the focusing of light along the longitudinal ($x$) and transverse ($y$) axes toward the intended focus site.

To obtain a smooth emission profile and a larger effective aperture, the grating strength was modified by linearly varying the duty cycle ($DC$) according to:
\begin{equation}
DC = DC_0 - R \sqrt{(x^2+y^2)},
\end{equation}
where the initial duty cycle, DC$_0$ and the rate at which the duty cycle was varied, $R$, were constrained by a combination of the minimum feature size and grating period obtained from the phase matching condition. To further reduce the minimum achievable grating strength, transverse magnetic (TM) polarization was chosen to minimize the mode overlap with the grating structure.

The grating design was optimized using two-dimensional finite-difference time-domain (2D-FDTD) simulations on the $y=0$ plane by adjusting the longitudinal location $x_0$ and $R$, while the focus height, $z_0$, was fixed at 50 {\textmu}m and the initial duty cycle $DC_0$ was maximized. Once these parameters were finalized, a 3D FDTD simulation was performed with the final structure to validate the grating design. The simulated light emission profiles in the $y=0$ plane from the 3D FDTD simulations are shown in Fig. \ref{fig:design} (d,e). 

A grating emitter for blue light ($\lambda = 488$ nm), targeting ChR-2 \cite{Lin2011} was designed and fabricated on 120nm thick plasma enhanced chemical vapour deposition (PECVD) SiN with  $x_0$ = 71.1 {\textmu}m and $R$ = 6003.6 m\textsuperscript{-1}. Another grating emitter for red light, targeting Chrimson \cite{Klapoetke2014} ($\lambda = 594$ nm) was designed and fabricated on 200nm thick PECVD SiN with the parameters $x_0$ = 75.5 {\textmu}m and $R$ = 4854.6 m\textsuperscript{-1}. Both emitter designs fit within an area of 20 {\textmu}m $\times$ 60 {\textmu}m to allow up to 16 such emitters on a single 100{\textmu}m-wide shank. 

Optimization of the grating parameters to maximize the numerical aperture of the focal point requires a balance between maintaining a wide range of emission angles and a uniform aperture. Because the grating design is also constrained by the minimum feature size of the fabrication process, the finalized designs contain grating periods that contribute to higher-order grating modes. 3D FDTD simulations shown in Fig. \ref{fig:design}(d,e) have a peak intensity ratio of -15.3 dB and -13.1 dB between the higher order grating mode and the focal spot for the blue and red emitter designs, respectively.

\subsection*{Fabrication}

\begin{figure}
    \centering
    \includegraphics[width=1\textwidth]{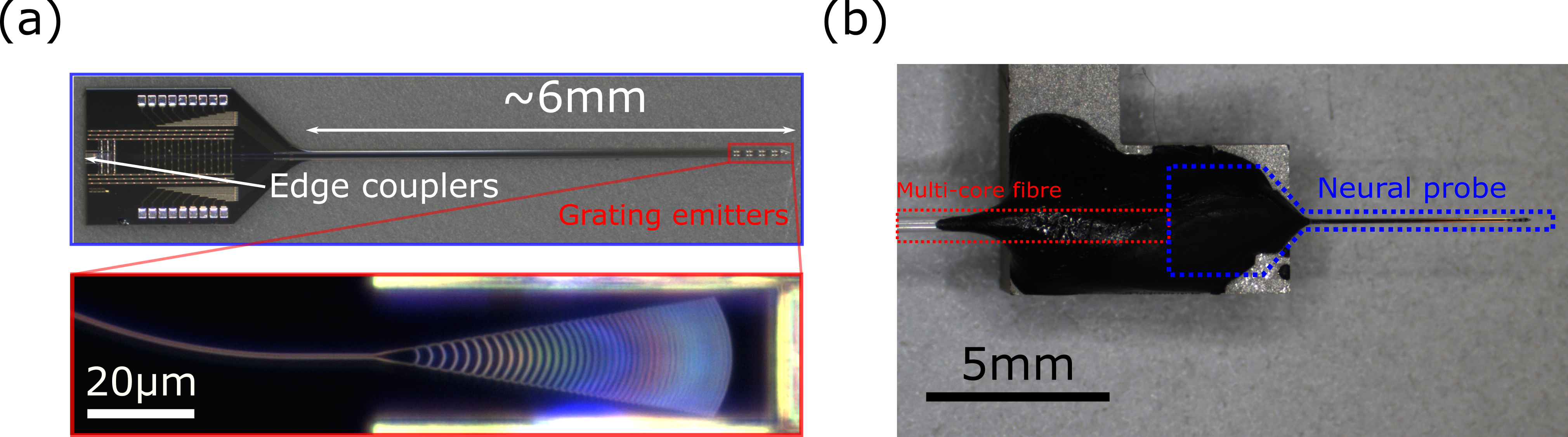}
            \caption{The Si neural probe with out-of-plane focusing grating emitters. (a) \textit{(top)}: Annotated micrograph of a neural probe. \textit{(bottom)}: Optical micrograph of an out-of-plane focusing grating emitter (brightness and contrast adjusted). (b) A photograph of a  neural probe attached to the input multicore fiber.}
            \label{fig:fab}
\end{figure}

The Si neural probes were fabricated on 200mm diameter Si wafers at Advanced Micro Foundry (AMF) using 193nm deep ultraviolet (DUV) lithography. The PECVD SiN waveguide layer was deposited with thicknesses of 120 or 200 nm on different variants of the neural probe. The aluminum metal routing layers and titanium nitride electrodes for electrophysiology recordings were available\cite{Chen2023} but not used in this work. The neural probe was defined with a deep trench etch, which was then released by thinning the Si substrate to $ \sim 100$ {\textmu}{m} with backgrinding. Figure \ref{fig:fab}(a) shows one of the fabricated probes. The cross-sectional area of the shank was $\sim$ 100{\textmu}m $\times$100{\textmu}m, comparable to other implantable probes \cite{Pisanello2018,Libbrecht2018}.

\subsection*{Packaging}

The Si neural probe was first fixed to a handle holder with a thermally curable epoxy. Then, it was packaged by aligning each edge coupler on the probe with a core in a multicore fiber and gluing it in place with a UV-curable epoxy using a custom semi-automatic machine (Ficontec) \cite{Azadeh2021}. Lastly, a black epoxy was manually applied over the UV-cured epoxy to block any stray light emission from the fiber-chip interface. Each emitter on the neural probe was spatially addressed with a micro-electromechanical system (MEMS) mirror system by coupling light into one of the 16 cores in the multicore fiber on the distal end using the configuration described in \cite{Chen2022, Chen2023}. This method of addressing the grating emitters allows the neural probe to be entirely passive to minimize heating in tissue.

\section*{Experiment and Results}

\begin{figure}
    \centering
    \includegraphics[width=1\textwidth]{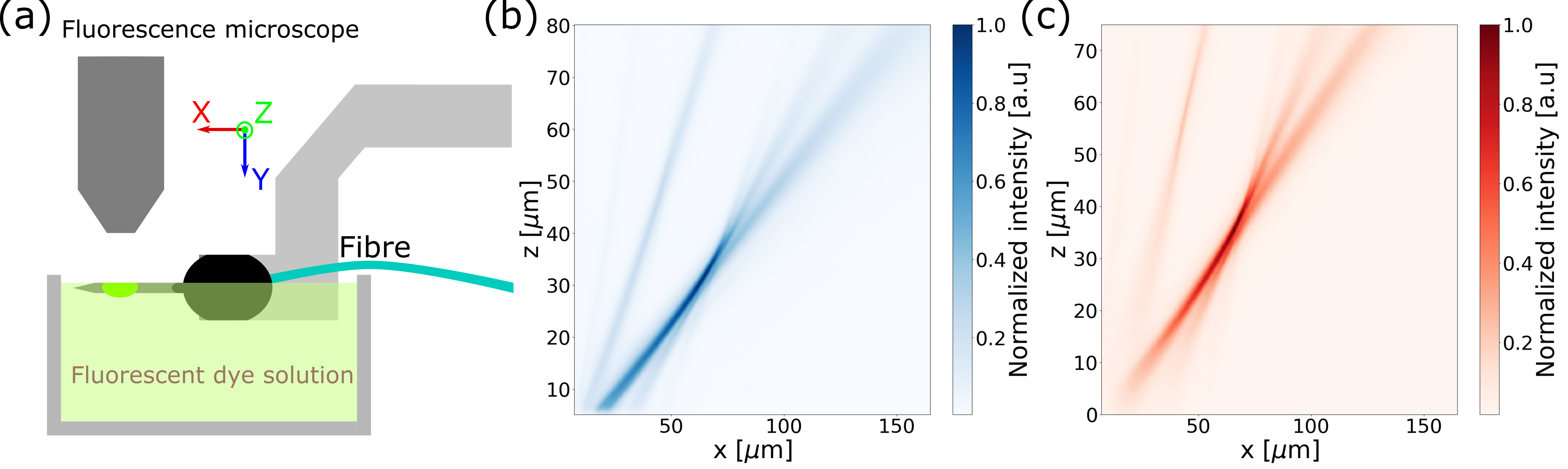}
            \caption{Side-view profile measurements by immersing the probe in a bath of fluorescent dye solution. (a) Diagram of measurement setup.  Captured side profile for (b)  $\lambda = 488$ nm, and (c) $\lambda = 594$ nm.}
            \label{fig:sideprof}
\end{figure}

The side-view of the emission profile of the grating emitters was captured by immersing the fiber-attached probe sideways in a mixture of water and fluorescent dye with a concentration of 100 {\textmu}M. The probe was oriented such that the captured optical beam profile was aligned with the $x-z$ plane. An illustration of this setup is shown in Fig. \ref{fig:sideprof}(a). Sodium fluorescein dye was used for $\lambda = 488$ nm and sulforhodamine 101 dye (Texas Red) was used for $\lambda = 594$ nm. The fluorescence was then captured with a microscope equipped with the suitable emission filter to isolate the fluorescent signal. However, since the emitted beam is focused in two dimensions, the side profile only captures a projection of the beam profile onto the $x-z$ plane. 

\begin{figure}
    \centering
    \includegraphics[width=0.8\textwidth]{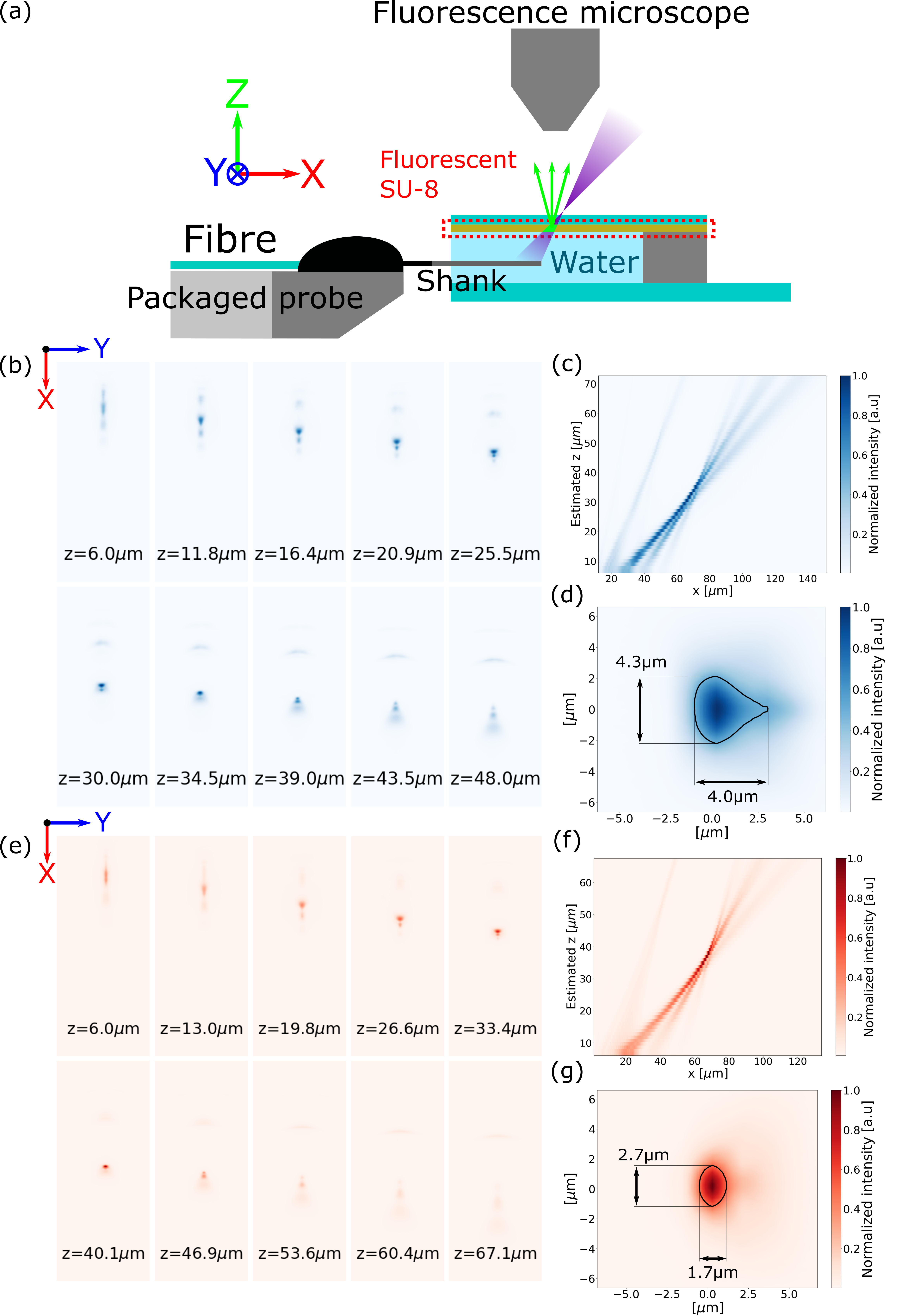}
            \caption{Fluorescent coverslip measurement to obtain the 3D volumetric emission pattern of the grating emitter. (a) Diagram of the measurement setup. (b) Captured cross-section profiles ($x-y$ plane) at various heights above the grating for $\lambda = 488$ nm. (c) Beam intensity on the $y=0$ plane obtained by stitching together the captured cross-sections for $\lambda = 488$ nm. (d) FWHM measurement of the interpolated beam waist for $\lambda = 488$ nm. (e, f, g): The corresponding images to (b, c, d) for $\lambda = 594$ nm.}
            \label{fig:coverslip}
\end{figure}

To characterize the full 3D beam profile, a fluorescent coverslip was made by spin coating a nominally 2{\textmu}m-thick layer of SU-8 photoresist mixed with sodium fluorescein or Texas Red dye using a procedure similar to that of Lim et al.\cite{Lim2018}. The fluorescent coverslip was used as the top plate (with the fluorescent side down) of a small chamber containing water to mimic the refractive index of brain tissue. The fiber-attached probe was then inserted into the water chamber and translated in the $z$ direction using a programmable micromanipulator in increments of $\sim$ 1 {\textmu}m. An illustration of this measurement is shown in Fig. \ref{fig:coverslip} (a).

The $x-y$ cross-sections of the beam, averaged over the thickness of the SU-8 layer, at various $z$ positions above the grating were captured by the fluorescent coverslip and imaged with the fluorescence microscope, and are shown in Fig. \ref{fig:coverslip} (b,e). From linecuts of the $x-y$ cross-sections, the side profiles of the beam on the $y=0$ plane were constructed and are shown in Fig. \ref{fig:coverslip} (c,f). Finally, the beam waist profiles interpolated from the 3D profiles are shown in Fig. \ref{fig:coverslip} (d,g). The widths (full-width at half-maximum (FWHM)) of the beam waists  were 4.0 $\mathrm{\mu m} \times 4.3$ $\mathrm{\mu m}$ and $1.7$ $\mathrm{\mu m} \times 2.7 $ $\mathrm{\mu m}$  for $\lambda = 488$ nm and $\lambda = 594$ nm, respectively. The ratio between the peak intensity of the unwanted grating order and that of the focal point was found to be -7 dB and -10.5 dB for the blue and red emitter designs, respectively.

\begin{figure}
    \centering
    \includegraphics[width=1\textwidth]{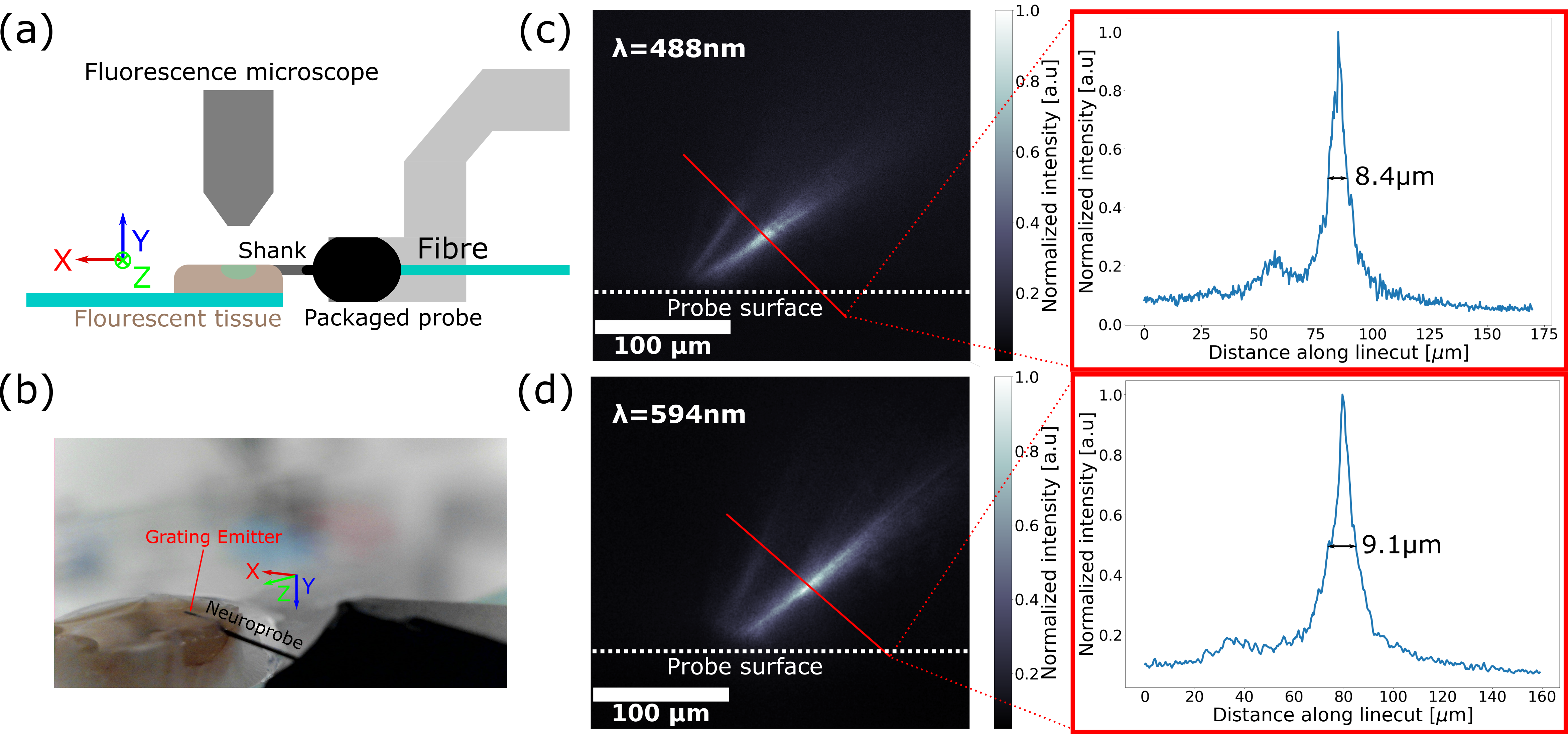}
            \caption{Verification of light focusing in fixed tissue. (a) Diagram of the experimental setup. (b) Photograph of implanted neural probe. Captured fluorescent side profiles with background subtracted ($x-z$ plane) of the emitted beam and linecut at the beam waist from (c) red emitter implanted in fixed tissue with jRGECO1a expression, and (d) blue emitter implanted in fixed tissue with GCaMP6s expression.}
            \label{fig:tissue}
\end{figure}

Finally, to validate optical focusing in tissue, the packaged probes were inserted into fixed brain tissue that expressed a genetically encoded calcium indicator (GECI). Tissues with Thy1-GCaMP6s expression were used for $\lambda = 488nm$ \cite{Dana2014} and tissue with Thy1-jRGECO1a expression was used for $\lambda = 594$ nm \cite{Dana2016}. The probe was inserted into the tissue in the same orientation as the side profile measurement, such that the captured profile was aligned to the $x-z$ plane and close to the surface of the tissue to obtain a profile that was minimally blurred by propagation through the tissue. The resulting fluorescent side profile was then captured with the fluorescence microscope. The experimental setup, the fluorescent side profiles of the beam focusing in tissue, and line cuts of the beam waists are shown in Fig. \ref{fig:tissue}. By measuring the beam waist from the side profiles, we estimate  the beam waist width (FWHM) in tissue to be 8.4 and 9.1 {\textmu}m for the blue and red emitter designs, respectively. 

\section*{Discussion}

\begin{figure}
    \centering
    \includegraphics[width=1\textwidth]{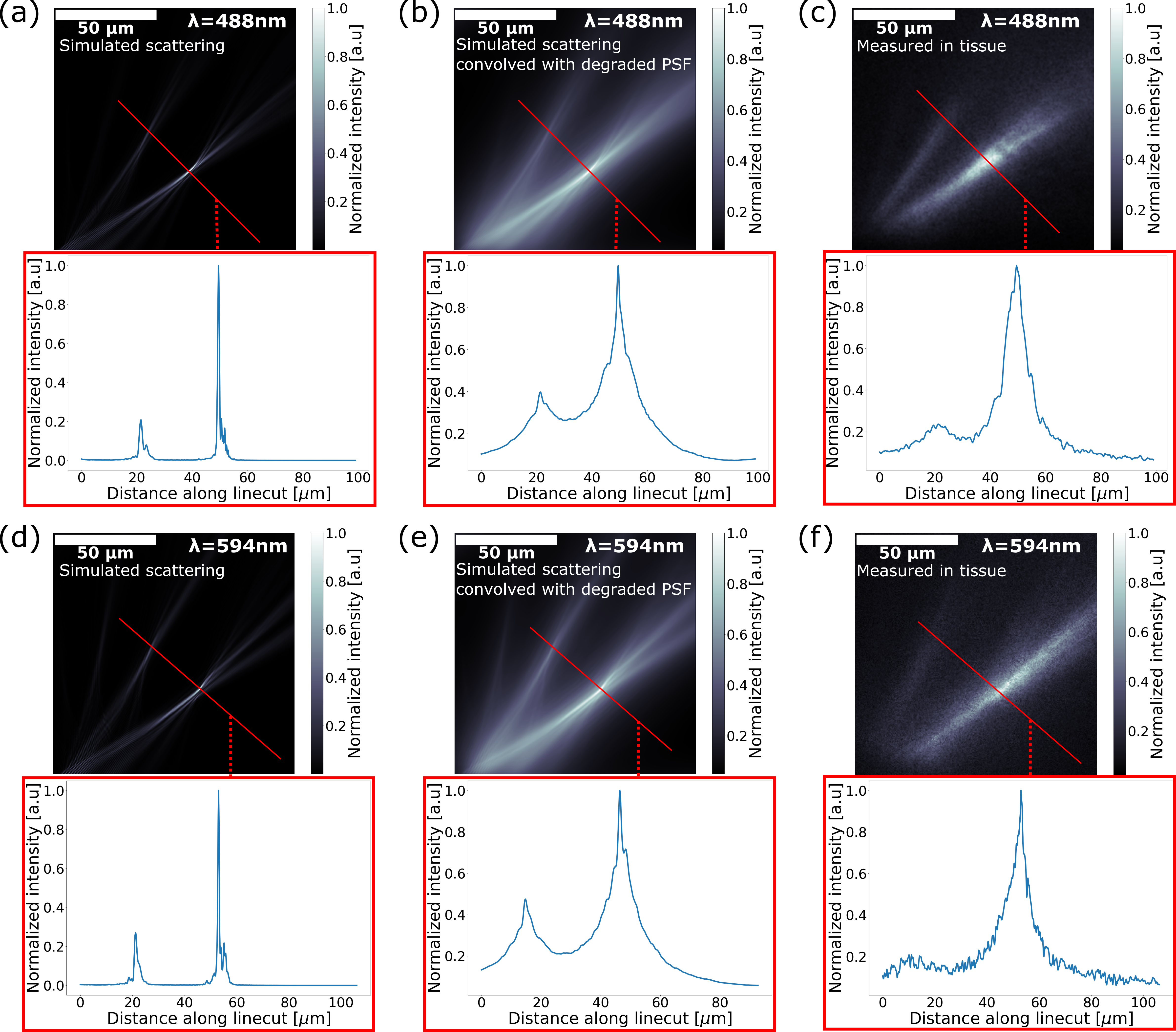}
            \caption{
            Comparison of the scattering simulation results with the measured beam profile in tissue. (a) Simulation of the blue emitter beam intensity profile in tissue using beam propagation with scattering for $\lambda = 488$ nm. (b) Simulated beam intensity side profile convolved with a degraded PSF obtained at $\lambda = 514$ nm. An implantation depth of 280 {\textmu}m was assumed. (c) Measured beam intensity side profile (background subtracted) in fixed tissue with GCaMP6s expression for the blue emitter design. (d) Simulation of the red emitter beam intensity profile in tissue using beam propagation with scattering for $\lambda = 594$ nm. (e) Simulated beam intensity side profile convolved with a degraded PSF obtained at $\lambda = 600$ nm. An implantation depth of 310 {\textmu}m was assumed. (f) Measured beam intensity side profile (background subtracted) in fixed tissue with jRGECO1a expression for the red emitter design.}
            \label{fig:beam_scatter}
\end{figure}

In the measurement of the beam emission in the fluorescent dye solution, the captured side profile ($x-z$ plane) compressed the $y$-axis of the beam profile. Because the higher-order grating emission was spread out over the $y$-axis, as can be seen in the captured cross-sections in Fig. \ref{fig:coverslip} (b,e), this caused the higher grating order to appear more prominent in Fig. \ref{fig:sideprof} (b,c).

In the tissue measurements, the neural probe was implanted as close to the surface as possible; however, it was difficult to predict or control the depth of the probe implantation. The measured emission side profile in tissue consisted of a combination of the scattering of the beam from the grating and the scattering of the fluorescent signal. These scattering effects led to a discrepancy between the side profiles imaged in the fluorescent dye solution in Fig. \ref{fig:sideprof} (b,c) and the side profiles imaged in tissue in Fig. \ref{fig:tissue} (c,d).

To simulate these scattering effects, we used a beam propagation scattering model with fractal refractive index variations using the method described by Glaser et al. \cite{Glaser2016}. Using the relationships  in Rogers et al. \cite{Rogers2009}, we tuned the parameters of the fractal model based on the power law dependence of the reduced scattering coefficient on wavelength $\mu^\prime_s (\lambda) \propto \lambda^{-1.127}$, the scattering coefficient $\mu_s = 170$cm$^{-1}$, and the absorption coefficient $\mu_a = 5$ cm$^{-1}$ \cite{Mesradi2013,Azimipour2014}.

We simulated the scattering from the grating into the tissue with complex fields captured from 3D FDTD simulations to generate a 3D intensity profile of the emitted beam in tissue. Fig. \ref{fig:beam_scatter} (a,c) shows a side view of the flattened intensity profile. This 3D intensity profile was then convolved with degraded point spread functions (PSFs) and flattened to emulate the image captured by the fluorescent microscope, as shown in Fig. \ref{fig:beam_scatter} (b,d). The degraded PSF was obtained by forward propagating with scattering a Gaussian approximation of the Airy disk defined by the $20\times$ infinity-corrected microscope objective \cite{Mitutoyo2022} at the peak emission wavelengths of the GECI ($\lambda = 514 nm$ for GCaMP6s and $\lambda = 600 nm$ for jRGECO1a) and  propagating in reverse without scattering. We find that the side profiles measured in tissue matched well with the simulation after taking into account the scattering, using our model, of the emitted beam and the fluorescent signal, assuming an implantation depth of 280 {\textmu}m and 310 {\textmu}m for the blue and red emitters, respectively. The discrepancy between the side profile obtained in the fluorescent dye solution and the tissue with GECI expression was dominated by the scattering of the fluorescent signal rather than the scattering of the beam emitted by the grating. This indicates that the measured FWHM of the focal spot in tissue overestimates the beam width in tissue.

Although the emitter focused as expected, the location of the focus deviated from the simulation. The observed focal height ($z_0$) and the beam uniformity over the $x$-axis were both lower than expected. This could suggest that the refractive index of the PECVD SiN in the fabricated device was higher than expected, which would have increased the  emission angle (to lower $z_0$) and increased the grating strength. A stronger grating would have made the emission along the $x$-axis less uniform and reduced the aperture dimension along the $x$ and $y$ axes due to the tapered design of the grating emitter. To reduce the dependence of the aperture size along the $y$ axis on the grating strength, future designs can widen the input waveguide instead of the grating emitter at the expense of a larger device footprint. Nevertheless, the measured focal spot size was comparable to the size of a neuronal soma \cite{Meitzen2011}. The spatial localization of light can be combined with optogenetic actuators that are targeted to express in specific structures of neurons \cite{Baker2016}.

With an incident power of 2 mW on the distal end of the fiber attached to the neural probe, the highest measured power of the blue emitter design was 4.5 {\textmu}W. Combining this with the beam waist extracted from the coverslip experiment, an average intensity of $\sim 80$ mW/mm\textsuperscript{2} is found within the contour region seen in Fig. \ref{fig:coverslip} (d), which is almost two orders of magnitude higher than the $\sim 1$ mW/mm\textsuperscript{2} threshold for optogenetic actuators \cite{Lin2011}. Thus, we expect that the probe could deliver sufficient optical intensities for optogenetic stimulation.

In summary, we have designed, fabricated, and characterized implantable neural probes with grating emitters that focus light out of the plane of the probe. In a non-scattering medium, the FWHM beam waists were 4.0 {\textmu}m $\times$ 4.3 {\textmu}m and 1.7 {\textmu}m $\times$ 2.7 {\textmu}m for the blue and red emitters, respectively. In fixed brain tissues with GECI expression, the scattering of the fluorescence signal led to broadened FWHM beams width of 8.4 and 9.1 {\textmu}m for the blue and red emitters, respectively. Although live tissue experiments were not performed, the probes delivered sufficient intensities for optogenetic stimulation. The generation of focused spots with a size scale of a neuronal soma in brain tissue using an implantable probe is promising for applications in spatially precise optogenetic experiments in deep brain regions. 

\bibliography{main}

\begin{thebibliography}{10}

\bibitem{Deisseroth2015}
Karl Deisseroth.
\newblock Optogenetics: 10 years of microbial opsins in neuroscience.
\newblock {\em Nature Neuroscience}, 18(9):1213--1225, Sep 2015.

\bibitem{Lin2011}
John~Y. Lin.
\newblock {A user's guide to channelrhodopsin variants: Features, limitations and future developments}.
\newblock {\em Experimental Physiology}, 96(1):19--25, 2011.

\bibitem{Klapoetke2014}
Nathan~C. Klapoetke, Yasunobu Murata, Sung~Soo Kim, Stefan~R. Pulver, Amanda Birdsey-Benson, Yong~Ku Cho, Tania~K. Morimoto, Amy~S. Chuong, Eric~J. Carpenter, Zhijian Tian, Jun Wang, Yinlong Xie, Zhixiang Yan, Yong Zhang, Brian~Y. Chow, Barbara Surek, Michael Melkonian, Vivek Jayaraman, Martha Constantine-Paton, Gane Ka-Shu Wong, and Edward~S. Boyden.
\newblock Independent optical excitation of distinct neural populations.
\newblock {\em Nature Methods}, 11(3):338--346, Mar 2014.

\bibitem{Al-Juboori2013}
Saif~I. Al-Juboori, Anna Dondzillo, Elizabeth~A. Stubblefield, Gidon Felsen, Tim~C. Lei, and Achim Klug.
\newblock {Light Scattering Properties Vary across Different Regions of the Adult Mouse Brain}.
\newblock {\em PLoS ONE}, 8(7):1--9, 2013.

\bibitem{Adesnik2021}
Hillel Adesnik and Lamiae Abdeladim.
\newblock Probing neural codes with two-photon holographic optogenetics.
\newblock {\em Nature Neuroscience}, 24(10):1356--1366, Oct 2021.

\bibitem{Neutens2023}
P~Neutens, J~O'Callaghan, J~De Ceulaer, E~Tonon, M~Welkenhuysen, C~M Lopez, A~Andrei, J~Putzeys, Md~Mahmud-Ul-Hasan, H~A~C Tilmans, and B~Dutta.
\newblock {Dual-wavelength neural probe for simultaneous opto-stimulation and recording fabricated in a monolithically integrated CMOS / photonics technology platform}.
\newblock In {\em International Electron Devices Meeting}, 2023.

\bibitem{Mohanty2020}
Aseema Mohanty, Qian Li, Mohammad~Amin Tadayon, Samantha~P. Roberts, Gaurang~R. Bhatt, Euijae Shim, Xingchen Ji, Jaime Cardenas, Steven~A. Miller, Adam Kepecs, and Michal Lipson.
\newblock Reconfigurable nanophotonic silicon probes for sub-millisecond deep-brain optical stimulation.
\newblock {\em Nature Biomedical Engineering}, 4(2):223--231, Feb 2020.

\bibitem{Pisanello2018}
Marco Pisanello, Filippo Pisano, Leonardo Sileo, Emanuela Maglie, Elisa Bellistri, Barbara Spagnolo, Gil Mandelbaum, Bernardo~L. Sabatini, Massimo De~Vittorio, and Ferruccio Pisanello.
\newblock Tailoring light delivery for optogenetics by modal demultiplexing in tapered optical fibers.
\newblock {\em Scientific Reports}, 8(1):4467, Mar 2018.

\bibitem{Sacher2019}
Wesley~D. Sacher, Xianshu Luo, Yisu Yang, Fu-Der Chen, Thomas Lordello, Jason C.~C. Mak, Xinyu Liu, Ting Hu, Tianyuan Xue, Patrick Guo-Qiang Lo, Michael~L. Roukes, and Joyce K.~S. Poon.
\newblock Visible-light silicon nitride waveguide devices and implantable neurophotonic probes on thinned 200 mm silicon wafers.
\newblock {\em Opt. Express}, 27(26):37400--37418, Dec 2019.

\bibitem{Sacher2021}
Wesley~D. Sacher, Fu-Der Chen, Homeira Moradi-Chameh, Xianshu Luo, Anton Fomenko, Prajay Shah, Thomas Lordello, Xinyu Liu, Ilan {Felts Almog}, John~N. Straguzzi, Trevor~M. Fowler, Youngho Jung, Ting Hu, Junho Jeong, Andres~M. Lozano, Patrick Guo-Qiang Lo, Taufik~A. Valiante, Laurent~C. Moreaux, Joyce K.~S. Poon, and Michael~L. Roukes.
\newblock {Implantable photonic neural probes for light-sheet fluorescence brain imaging}.
\newblock {\em Neurophotonics}, 8(02):1--26, 2021.

\bibitem{Sacher2022}
Wesley~D. Sacher, Fu-Der Chen, Homeira Moradi-Chameh, Xinyu Liu, Ilan~Felts Almog, Thomas Lordello, Michael Chang, Azadeh Naderian, Trevor~M. Fowler, Eran Segev, Tianyuan Xue, Sara Mahallati, Taufik~A. Valiante, Laurent~C. Moreaux, Joyce K.~S. Poon, and Michael~L. Roukes.
\newblock Optical phased array neural probes for beam-steering in brain tissue.
\newblock {\em Opt. Lett.}, 47(5):1073--1076, Mar 2022.

\bibitem{Chen2023}
Fu~Der Chen, Homeira~Moradi Chameh, Mandana Movahed, Hannes Wahn, Xin Mu, Peisheng Ding, Tianyuan Xue, John~N. Straguzzi, David~A. Roszko, Ankita Sharma, Alperen Govdeli, Youngho Jung, Hongyao Chua, Xianshu Luo, Patrick G.~Q. Lo, Taufik~A. Valiante, Wesley~D. Sacher, and Joyce K.~S. Poon.
\newblock Implantable nanophotonic neural probes for integrated patterned photostimulation and electrophysiology recording.
\newblock {\em bioRxiv}, 2023.

\bibitem{Voroslakos2022}
Mihály Vöröslakos, Kanghwan Kim, Nathan Slager, Eunah Ko, Sungjin Oh, Saman~S. Parizi, Blake Hendrix, John~P. Seymour, Kensall~D. Wise, György Buzsáki, Antonio Fernández-Ruiz, and Euisik Yoon.
\newblock Hectostar {\textmu}led optoelectrodes for large-scale, high-precision in vivo opto-electrophysiology.
\newblock {\em Advanced Science}, 9(18):2105414, 2022.

\bibitem{Schwaerzle2017}
M~Schwaerzle, O~Paul, and P~Ruther.
\newblock Compact silicon-based optrode with integrated laser diode chips, su-8 waveguides and platinum electrodes for optogenetic applications.
\newblock {\em Journal of Micromechanics and Microengineering}, 27(6):065004, apr 2017.

\bibitem{Libbrecht2018}
Sarah Libbrecht, Luis Hoffman, Marleen Welkenhuysen, Chris {Van den Haute}, Veerle Baekelandt, Dries Braeken, and Sebastian Haesler.
\newblock {Proximal and distal modulation of neural activity by spatially confined optogenetic activation with an integrated high-density optoelectrode}.
\newblock {\em Journal of Neurophysiology}, 120(1):149--161, 2018.

\bibitem{Kim2013}
Tae il~Kim, Jordan~G. McCall, Yei~Hwan Jung, Xian Huang, Edward~R. Siuda, Yuhang Li, Jizhou Song, Young~Min Song, Hsuan~An Pao, Rak-Hwan Kim, Chaofeng Lu, Sung~Dan Lee, Il-Sun Song, GunChul Shin, Ream Al-Hasani, Stanley Kim, Meng~Peun Tan, Yonggang Huang, Fiorenzo~G. Omenetto, John~A. Rogers, and Michael~R. Bruchas.
\newblock Injectable, cellular-scale optoelectronics with applications for wireless optogenetics.
\newblock {\em Science}, 340(6129):211--216, 2013.

\bibitem{Yasunaga2022}
Hiroki Yasunaga, Hibiki Takeuchi, Koyo Mizuguchi, Atsushi Nishikawa, Alexander Loesing, Mikiko Ishikawa, Chikako Kamiyoshihara, Susumu Setogawa, Noriaki Ohkawa, and Hiroto Sekiguchi.
\newblock Microled neural probe for effective in vivo optogenetic stimulation.
\newblock {\em Opt. Express}, 30(22):40292--40305, Oct 2022.

\bibitem{Mu2023}
Xin Mu, Fu-Der Chen, Ka~My Dang, Michael G.~K. Brunk, Jianfeng Li, Hannes Wahn, Andrei Stalmashonak, Peisheng Ding, Xianshu Luo, Hongyao Chua, Guo-Qiang Lo, Joyce K.~S. Poon, and Wesley~D. Sacher.
\newblock Implantable photonic neural probes with 3d-printed microfluidics and applications to uncaging.
\newblock {\em Frontiers in Neuroscience}, 17, 2023.

\bibitem{Chen2021}
Fu-Der Chen, Youngho Jung, Tianyuan Xue, Jason C.~C. Mak, Xianshu Luo, Patrick Guo-Qiang Lo, Michael~L. Roukes, Joyce K.~S. Poon, and Wesley~D. Sacher.
\newblock Sidelobe-free beam-steering using optical phased arrays for neural probes.
\newblock In {\em Conference on Lasers and Electro-Optics}, page SW3B.2. Optica Publishing Group, 2021.

\bibitem{Sharma2023}
Ankita Sharma, Alperen Govdeli, Tianyuan Xue, Fu-Der Chen, Xianshu Luo, Hongyao Chua, Guo-Qiang Lo, Wesley~D. Sacher, and Joyce~K.S. Poon.
\newblock Wide-angle single-lobe beam-steering using optical phased arrays on implantable neural probes.
\newblock In {\em CLEO 2023}, page SF2E.5. Optica Publishing Group, 2023.

\bibitem{Fiáth2019}
Rich{\'a}rd Fi{\'a}th, Adrienn~Lilla M{\'a}rton, Ferenc M{\'a}ty{\'a}s, Domonkos Pinke, Gergely M{\'a}rton, Kinga T{\'o}th, and Istv{\'a}n Ulbert.
\newblock Slow insertion of silicon probes improves the quality of acute neuronal recordings.
\newblock {\em Scientific Reports}, 9(1):111, Jan 2019.

\bibitem{Mehta2017}
Karan~K. Mehta and Rajeev~J. Ram.
\newblock Precise and diffraction-limited waveguide-to-free-space focusing gratings.
\newblock {\em Scientific Reports}, 7(1):2019, May 2017.

\bibitem{Corsetti2023}
Sabrina Corsetti, Ashton Hattori, Reuel Swint, Milica Notaros, Gavin~N. West, Tal Sneh, Felix Knollmann, Patrick~T. Callahan, Thomas Mahony, Ethan~R. Clements, Dave Kharas, Cheryl Sorace-Agaskar, Robert McConnell, John Chiaverini, and Jelena Notaros.
\newblock Integrated polarization-diverse grating emitters for trapped-ion quantum systems.
\newblock In {\em Frontiers in Optics $+$ Laser Science 2023 (FiO, LS)}, page JTu7A.3. Optica Publishing Group, 2023.

\bibitem{Becker2020}
Hanna Becker, Clemens~J. Krückel, Dries Van~Thourhout, and Martijn J.~R. Heck.
\newblock Out-of-plane focusing grating couplers for silicon photonics integration with optical mram technology.
\newblock {\em IEEE Journal of Selected Topics in Quantum Electronics}, 26(2):1--8, 2020.

\bibitem{Lanzio2018}
Vittorino Lanzio, Melanie West, Alexander Koshelev, Gregory Telian, Paolo Micheletti, Raquel Lambert, Scott Dhuey, Hillel Adesnik, Simone Sassolini, and Stefano Cabrini.
\newblock {High-density electrical and optical probes for neural readout and light focusing in deep brain tissue}.
\newblock {\em Journal of Micro/Nanolithography, MEMS, and MOEMS}, 17(2):025503, 2018.

\bibitem{Oton2016}
C.~J. Oton.
\newblock Long-working-distance grating coupler for integrated optical devices.
\newblock {\em IEEE Photonics Journal}, 8(1):1--8, 2016.

\bibitem{Azadeh2021}
Saeed~Sharif Azadeh, Andrei Stalmashonak, Kevin~W. Bennett, Fu-Der Chen, Wesley~D. Sacher, and Joyce K.~S. Poon.
\newblock Visible spectrum multicore fibers with 10 and 16 cores.
\newblock In {\em Conference on Lasers and Electro-Optics}, page STu4A.3. Optica Publishing Group, 2021.

\bibitem{Chen2022}
Fu-Der Chen, Hannes Wahn, Tianyuan Xue, Youngho Jung, John~N. Straguzzi, Saeed~S. Azadeh, Andrei Stalmashonak, Hongyao Chua, Xianshu Luo, Prajay Shah, Homeira~Moradi Chameh, Patrick Guo-Qiang Lo, Taufik~A. Valiante, Wesley~D. Sacher, and Joyce K.~S. Poon.
\newblock Implantable neural probe system for patterned photostimulation and electrophysiology recording.
\newblock In {\em Conference on Lasers and Electro-Optics}, page JTh6A.7. Optica Publishing Group, 2022.

\bibitem{Lim2018}
Miles~P. Lim, Xiaofei Guo, Eva~L. Grunblatt, Garrett~M. Clifton, Ayda~N. Gonzalez, and Christopher~N. LaFratta.
\newblock Augmenting mask-based lithography with direct laser writing to increase resolution and speed.
\newblock {\em Opt. Express}, 26(6):7085--7090, Mar 2018.

\bibitem{Dana2014}
Hod Dana, Tsai~Wen Chen, Amy Hu, Brenda~C. Shields, Caiying Guo, Loren~L. Looger, Douglas~S. Kim, and Karel Svoboda.
\newblock {Thy1-GCaMP6 transgenic mice for neuronal population imaging in vivo}.
\newblock {\em PLoS ONE}, 9(9), 2014.

\bibitem{Dana2016}
Hod Dana, Boaz Mohar, Yi~Sun, Sujatha Narayan, Andrew Gordus, Jeremy~P Hasseman, Getahun Tsegaye, Graham~T Holt, Amy Hu, Deepika Walpita, Ronak Patel, John~J Macklin, Cornelia~I Bargmann, Misha~B Ahrens, Eric~R Schreiter, Vivek Jayaraman, Loren~L Looger, Karel Svoboda, and Douglas~S Kim.
\newblock Sensitive red protein calcium indicators for imaging neural activity.
\newblock {\em eLife}, 5:e12727, mar 2016.

\bibitem{Glaser2016}
Adam~K. Glaser, Ye~Chen, and Jonathan T.~C. Liu.
\newblock Fractal propagation method enables realistic optical microscopy simulations in biological tissues.
\newblock {\em Optica}, 3(8):861--869, Aug 2016.

\bibitem{Rogers2009}
Jeremy~D. Rogers, \.{I}lker R.~\c{C}apo\u{g}lu, and Vadim Backman.
\newblock Nonscalar elastic light scattering from continuous random media in the born approximation.
\newblock {\em Opt. Lett.}, 34(12):1891--1893, Jun 2009.

\bibitem{Mesradi2013}
Mohammed Mesradi, Aurelie Genoux, Vesna Cuplov, Darine Abi-Haidar, Sebastien Jan, Irene Buvat, and Frederic Pain.
\newblock {Experimental and analytical comparative study of optical coefficient of fresh and frozen rat tissues}.
\newblock {\em Journal of Biomedical Optics}, 18(11):117010, 2013.

\bibitem{Azimipour2014}
Mehdi Azimipour, Ryan Baumgartner, Yuming Liu, Steven~L. Jacques, Kevin~W. Eliceiri, and Ramin Pashaie.
\newblock {Extraction of optical properties and prediction of light distribution in rat brain tissue}.
\newblock {\em Journal of Biomedical Optics}, 19(7):075001, 2014.

\bibitem{Mitutoyo2022}
{Microscopes Units and Objectives (UV, NUV, Visible \& NIR Region)}.
\newblock Technical report, Mitutoyo, 2022.

\bibitem{Meitzen2011}
John Meitzen, Kelsey~R. Pflepsen, Christopher~M. Stern, Robert~L. Meisel, and Paul~G. Mermelstein.
\newblock Measurements of neuron soma size and density in rat dorsal striatum, nucleus accumbens core and nucleus accumbens shell: Differences between striatal region and brain hemisphere, but not sex.
\newblock {\em Neuroscience Letters}, 487(2):177--181, 2011.

\bibitem{Baker2016}
Christopher~A Baker, Yishai~M Elyada, Andres Parra, and M~McLean Bolton.
\newblock Cellular resolution circuit mapping with temporal-focused excitation of soma-targeted channelrhodopsin.
\newblock {\em eLife}, 5:e14193, aug 2016.

\end{thebibliography}

\end{document}